\documentclass{article}
\usepackage{ijcai26}

\usepackage{times}
\usepackage{soul}
\usepackage{url}
\usepackage[hidelinks]{hyperref}
\usepackage{bookmark} 
\usepackage[utf8]{inputenc}
\usepackage[small]{caption}
\usepackage{graphicx}
\usepackage{amsmath}
\usepackage{amssymb}
\usepackage{amsfonts}
\usepackage{amsthm}
\usepackage{booktabs}
\usepackage{algorithm}
\usepackage{algorithmic}
\usepackage{threeparttable}
\usepackage{multirow}
\usepackage{natbib}
\usepackage{subcaption}
\usepackage{graphicx}

\title{HiDE: Hierarchical Dictionary-Based Entropy Modeling for Learned Image Compression}

\author{
Haoxuan Xiong\textsuperscript{1}
\and
Yuanyuan Xu\textsuperscript{1*}\and
Kun Zhu\textsuperscript{2}\and
Yiming Wang\textsuperscript{3}\and 
Baoliu Ye\textsuperscript{4}\\
\affiliations
\textsuperscript{1}Hohai University\\
\textsuperscript{2}Nanjing University of Aeronautics and Astronautics\\
\textsuperscript{3}Nanjing Audit University\\
\textsuperscript{4}Nanjing University\\
\emails
\{haoxuan\_x, yuanyuan\_xu\}@hhu.edu.cn,
zhukun@nuaa.edu.cn,
wangyiming@nau.edu.cn,
yebl@nju.edu.cn,
}

\begin{document}

\maketitle

\begin{abstract}

Learned image compression (LIC) has achieved remarkable coding efficiency, where entropy modeling plays a pivotal role in minimizing bitrate through informative priors. Existing methods predominantly exploit internal contexts within the input image, yet the rich external priors embedded in large-scale training data remain largely underutilized. Recent advances in dictionary-based entropy models have demonstrated that incorporating external priors can substantially enhance compression performance. However, current approaches organize heterogeneous external priors within a single-level dictionary, resulting in imbalanced utilization and limited representational capacity. Moreover, effective entropy modeling requires not only expressive priors but also a parameter estimation network capable of interpreting them. To address these challenges, we propose \textbf{HiDE}, a \textbf{Hi}erarchical \textbf{D}ictionary-based \textbf{E}ntropy modeling framework for learned image compression. HiDE decomposes external priors into global structural and local detail dictionaries with cascaded retrieval, enabling structured and efficient utilization of external information. Moreover, a context-aware parameter estimator with parallel multi-receptive-field design is introduced to adaptively exploit heterogeneous contexts for accurate conditional probability estimation. Experimental results show that HiDE achieves \textbf{18.5\%}, \textbf{21.99\%}, and \textbf{24.01\%} BD-rate savings over VTM-12.1 on the Kodak, CLIC, and Tecnick datasets, respectively.
\end{abstract}

\section{Introduction}

Image compression remains a fundamental challenge in multimedia communication, aiming to reduce storage and transmission costs while preserving reconstruction quality.
Recent advances in learned image compression (LIC) have outperformed traditional standards such as JPEG \citep{JPEG} and VVC-Intra \citep{VVC2021} in terms of rate-distortion (RD) performance.
LIC follow a variational autoencoder framework, where images are transformed into latent representations that are quantized and subsequently entropy coded. Within this framework, the entropy model is decisive for compression efficiency, as it models the probability distribution of the latent representation, which determines the bitrate through entropy coding.
\begin{figure}[t!]
  \centering
  \includegraphics[width=\linewidth]{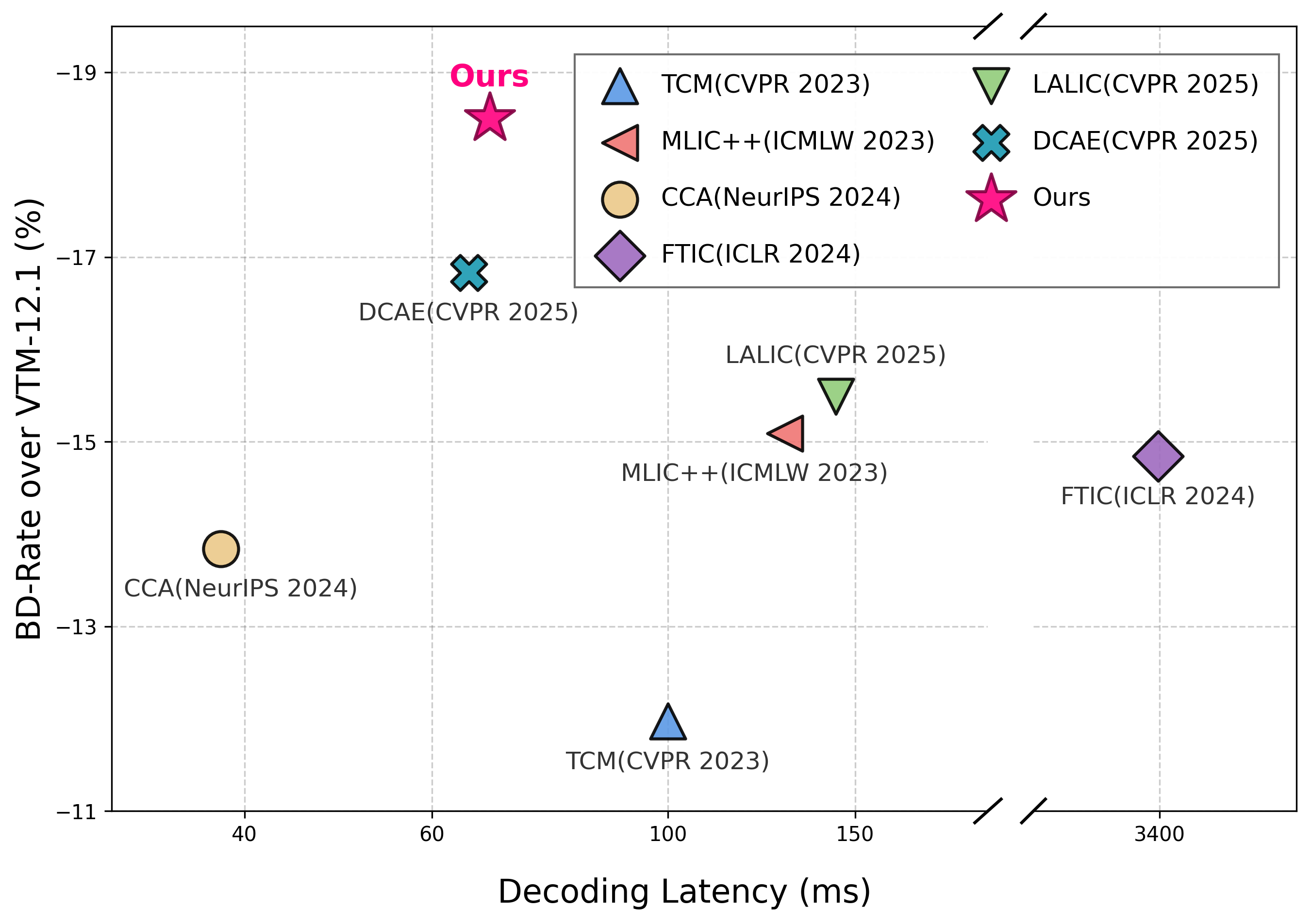}
  \setlength{\abovecaptionskip}{-4mm}
  \caption{BD-Rate vs. decoding latency on the Kodak dataset (top-left indicates better performance).}
  \label{fig:rate_speed}
\end{figure}

\begin{figure*}[t!] 
  \centering
  \setlength{\abovecaptionskip}{3pt} 
  \setlength{\belowcaptionskip}{-10pt} 


  \begin{subfigure}[b]{0.495\linewidth}
      \centering
      \includegraphics[width=\linewidth]{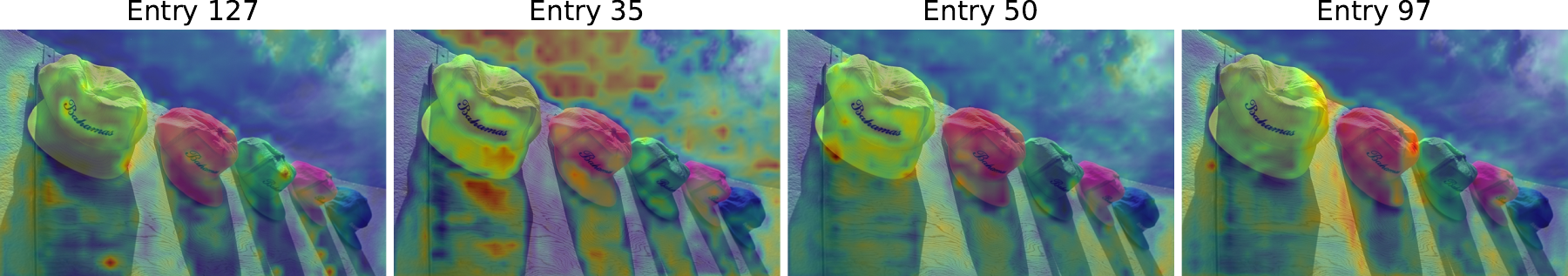} 
  \end{subfigure}
  \hfill
  \begin{subfigure}[b]{0.495\linewidth}
      \centering
      \includegraphics[width=\linewidth]{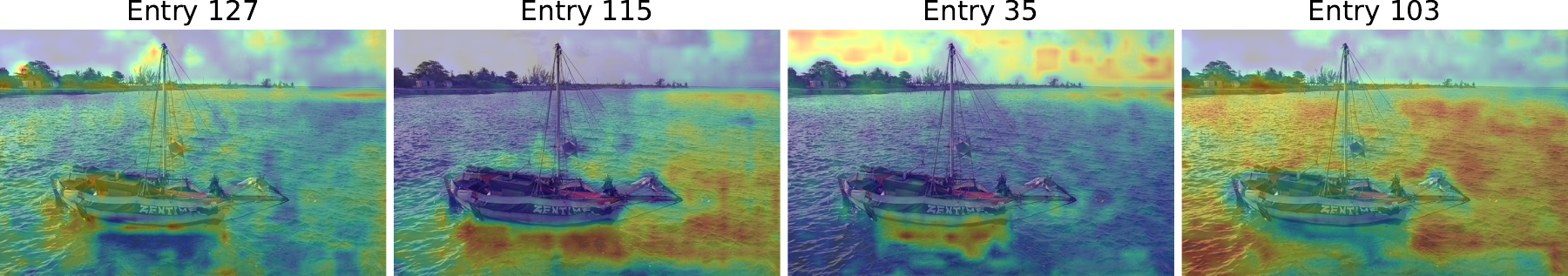}
  \end{subfigure}
  
  
  \begin{subfigure}[b]{0.495\linewidth}
      \centering
      \includegraphics[width=\linewidth]{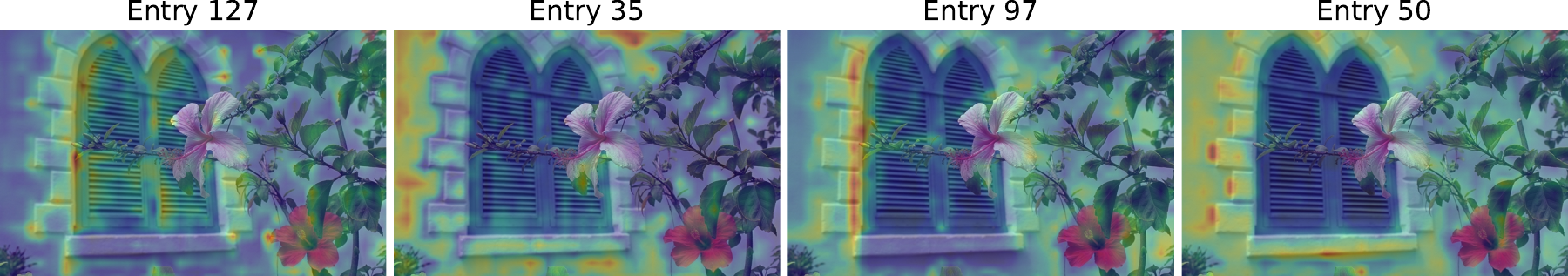}
  \end{subfigure}
  \hfill
  \begin{subfigure}[b]{0.495\linewidth}
      \centering
      \includegraphics[width=\linewidth]{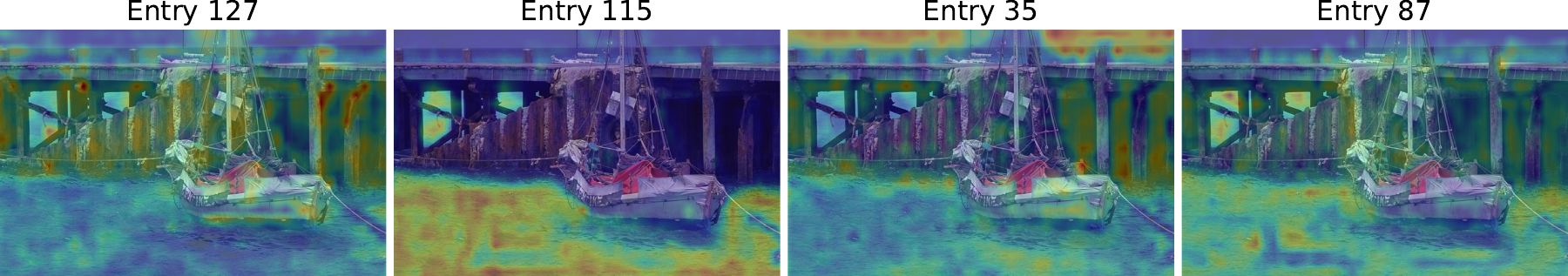}
  \end{subfigure}
  
  
  \begin{subfigure}[b]{0.33\linewidth}
      \centering
      \includegraphics[width=\linewidth]{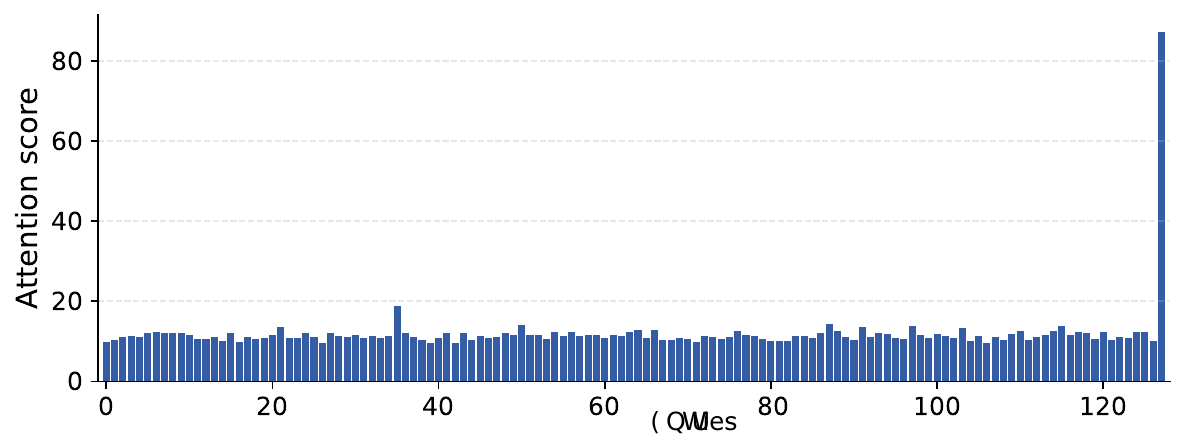}
      
  \end{subfigure}
  \hfill
  \begin{subfigure}[b]{0.33\linewidth}
      \centering
      \includegraphics[width=\linewidth]{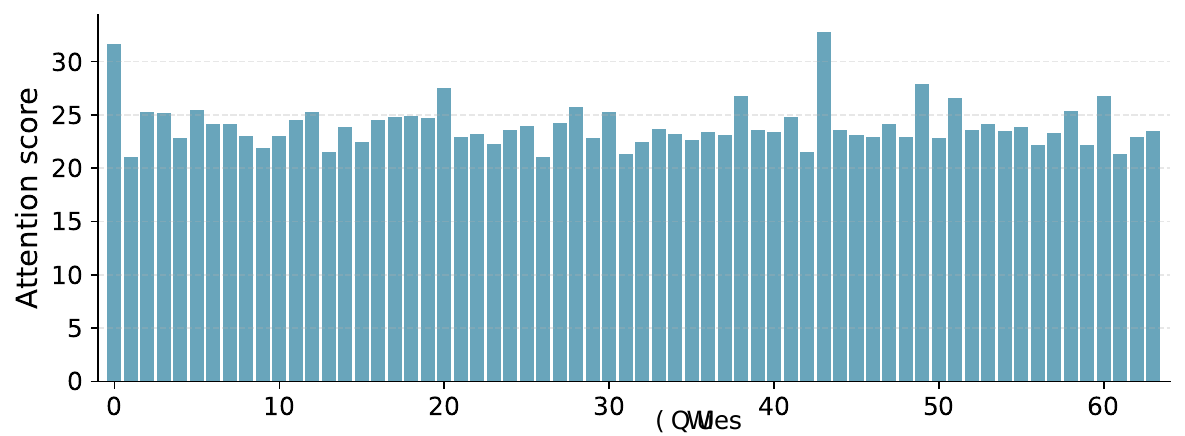}
      
  \end{subfigure}
  \hfill
  \begin{subfigure}[b]{0.33\linewidth}
      \centering
      \includegraphics[width=\linewidth]{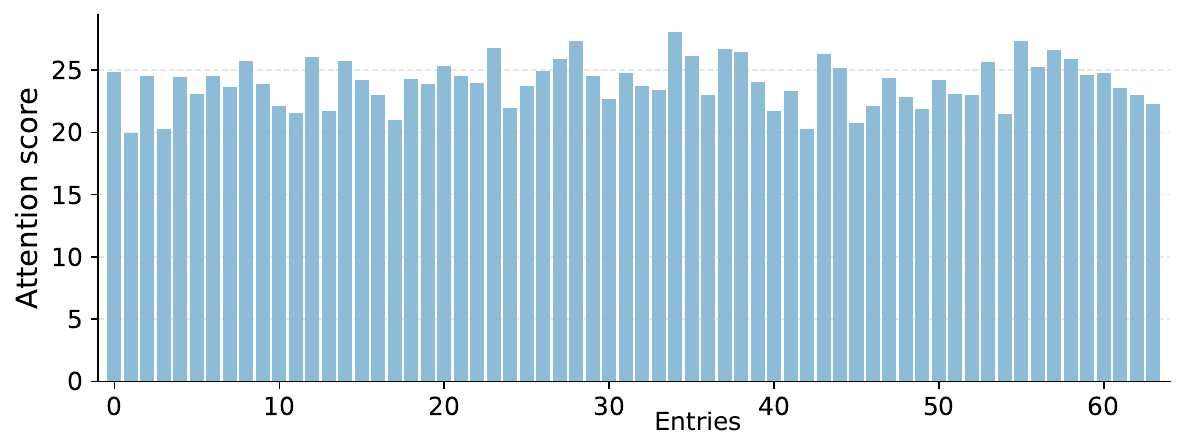}
      
  \end{subfigure}
  
  \caption{Visualization of dictionary entry utilization. 
The top two rows show attention heatmaps from DCAE on Kodak images, highlighting the dominant entries in the retrieval process. 
The bottom row presents the mean attention scores of dictionary entries across the Kodak dataset, where DCAE’s single-level dictionary (left) exhibits a highly skewed usage, while our proposed \textbf{Global} (middle) and \textbf{Detail} (right) dictionaries achieve more balanced utilization.}
  \label{fig:entry_usage}
\end{figure*}
To minimize the bitrate, it is essential to reduce the uncertainty of the latent representation.
Entropy models achieve this by conditioning probability estimation on informative priors, thereby reducing the conditional entropy of each latent symbol.
Seminal works~\citep{Balle17, Minnen_18} introduced hyperpriors to capture spatial dependencies and autoregressive models to leverage causal context. 
Subsequent research has focused on enhancing the expressiveness of context models to more effectively capture correlations among latent symbols. 
Approaches have evolved from channel-sliced~\citep{MinnenS20} and checkerboard mechanisms~\citep{He_21} to intricate spatial-channel context~\citep{He_22, Jiang23, jiang2023mlic}. Notably, the MLIC series~\citep{Jiang23, jiang2023mlic} employ advanced attention mechanisms that jointly integrate local, global, and channel-wise contexts, achieving impressive performance. 
Nevertheless, these methods rely exclusively on the internal context derived  from the input and overlook the rich statistical patterns embedded in large-scale training data.

Recently, the dictionary-based cross-attention entropy model (DCAE)~\citep{Lu2025} addressed this limitation by introducing a learned dictionary as an external prior. DCAE leverages cross-attention to retrieve relevant dictionary entries that serve as external priors for entropy modeling, thereby reducing the uncertainty of latent symbols. Even when combined with simple channel-wise context, DCAE achieves state-of-the-art performance, underscoring the effectiveness of external priors compared to increasingly complex internal context modeling.

However, representing diverse visual content using a finite dictionary is inherently prone to representation collapse, a well-known issue in vector quantized generative models~\citep{vqvae17,vqgan21, Collapse25}. In such cases, a few dictionary entries are frequently selected while the majority remain rarely utilized. To further investigate this phenomenon, we analyzed the dictionary utilization patterns in DCAE. As illustrated in Figure~\ref{fig:entry_usage}, the attention maps of dictionary entries exhibit a clear winner-takes-all tendency, where a small subset of generic patterns disproportionately dominates the retrieval process across diverse images. Specifically, Entry~127 consistently responds to complex structural regions, while Entries~35 and~115 primarily activate on smoother areas. The histogram in the bottom-left of Figure~\ref{fig:entry_usage} shows a markedly skewed distribution, where most dictionary entries receive relatively low attention scores, whereas a few entries exhibit significantly higher activations. This imbalance indicates that the external prior is unevenly exploited, degrading into a static bias rather than functioning as a dynamic, content-adaptive reference, thereby imposing a representational bottleneck on entropy modeling.

Furthermore, the availability of rich priors alone does not guarantee accurate conditional probability estimation. Effective entropy modeling requires a parameter estimation network capable of transforming heterogeneous context priors into appropriate entropy parameters. Existing approaches~\citep{MinnenS20, Liu23, Lu2025} typically adopt shallow convolutional estimators with fixed receptive fields to integrate hyperpriors, autoregressive contexts, and dictionary-based priors. As the diversity of contextual information increases, such limited architectures restrict the effective exploitation of richer priors and constrain model performance.

To address the aforementioned challenges and fully exploit the potential of external prior modeling, we propose \textbf{HiDE}, a hierarchical dictionary-based entropy model equipped with a context-aware parameter estimation network for learned image compression. Our main contributions are summarized as follows: 

\begin{itemize}
  \item We propose a hierarchical dictionary-based entropy framework that decomposes external priors into global structural and local detail dictionaries, facilitating structured and efficient utilization of external information.
  
  \item We design a context-aware parameter estimation network featuring multiple receptive fields context extractor, enabling adaptive exploitation of diverse contexts for more accurate conditional probability estimation.
  
  \item Comprehensive experiments demonstrate that HiDE consistently outperforms existing state-of-the-art methods on various benchmark datasets with competitive decoding speed, as evidenced in Figure~\ref{fig:rate_speed} and Table~\ref{tab:sota_comparison}.
\end{itemize}



\section{Related Work}
\subsection{Learned Image Compression}
\label{sec:related_lic}

LIC adopts a nonlinear transform framework~\citep{Balle17,Balle21}, typically parameterized by convolutional neural networks (CNNs), vision transformers (ViTs), or hybrid architectures combining the two~\citep{Cheng20,Zhong20,Zhu22,Liu23,Li_ftic_24,Lalic_25,zeng2025mambaic}. These studies primarily focus on architectural refinements to obtain more compact latent representations. However, performance improvements driven solely by network architecture are ultimately constrained by the representational capacity of the latent space.

To minimize coding cost, entropy models aim to reduce the cross-entropy between the predicted probability distribution and the actual distribution of the latents. The hyperprior model~\citep{Balle2018} introduces side information to estimate conditional entropy and improve probability modeling. Building on this, the autoregressive model~\citep{Minnen_18} leverages previously decoded elements as causal context, significantly enhancing predictive accuracy. To balance compression efficiency and decoding parallelism, subsequent methods proposed channel-sliced~\citep{MinnenS20} and checkerboard~\citep{He_21} context models, which partition latents into groups for parallel processing. 

More recent efforts have further enriched context modeling through multi-reference priors~\citep{He_22,Jiang23,jiang2023mlic} or channel-wise causal adjustment losses~\citep{Han24}. For instance, \citep{Jiang23} introduced intricate spatial-channel context modeling to capture inter- and intra-component correlations in the latent space. Despite these advances, these context models rely exclusively on internal information derived from the input, overlooking the rich external priors inherent in large-scale training data that could further enhance entropy modeling.

\subsection{Dictionary Learning}
\label{sec:related_dict}

Dictionary learning provides an effective paradigm for exploiting external data priors. In generative modeling, vector quantization (VQ) methods~\citep{vqvae17, vqgan21} demonstrate that learned dictionaries can summarize complex visual patterns by representing images as compositions of discrete code entries. 

In the context of image compression, early approaches~\citep{MinnenTSHC18} utilized a static and non-learnable dictionary, which lacked the flexibility to adapt to diverse external knowledge. More recently, Conditional Latent Coding (CLC) \citep{WuCLH25} constructs a feature dictionary to provide conditional references for latent adjustment, while Mask-based Selective Compensation (MSC) \citep{Kuang_2025_ICCV} retrieves compensation vectors to correct residual errors. Although these methods validate the benefit of external repositories, they focus primarily on reconstruction rather than entropy estimation.

Dictionary-based cross attention entropy model (DCAE)~\citep{Lu2025} addressed this limitation by integrating dictionary priors with internal priors to improve probability estimation. 
Despite its effectiveness, DCAE employs a single-level dictionary to represent all visual patterns. As noted in our analysis, this flat design is prone to representation collapse, which causes unbalanced dictionary utilization and hampers model expressiveness.

\subsection{Entropy Parameter Estimation}
\label{sec:related_estimator}

While context models capture dependencies among latent variables, the parameter estimation network plays a crucial role in mapping these priors to the parameters of the conditional latent distribution. The Gaussian Scale Mixture (GSM) model~\citep{Balle2018} predicts the scale of latent variables, whereas \citep{Minnen_18} extended GSM to jointly estimate both mean and scale, achieving more accurate density modeling. Latent Residual Prediction (LRP)~\citep{MinnenS20} further refines quantization error prediction. These works established the practice of predicting the mean, scale, and potentially the residual for entropy coding.

Despite the increasing sophistication of context models, the architecture of parameter estimation networks has remained largely unchanged. Most approaches employ shallow convolutional estimators with fixed receptive fields, regardless of the heterogeneity of input priors. However, modern entropy models integrate highly heterogeneous context sources, including hyperpriors encoding global statistics, autoregressive context capturing local causal dependencies, and more recently, external dictionary-based priors~\citep{Balle2018, MinnenTSHC18, MinnenS20, He_21,He_22, Jiang23, Lu2025}. Applying fixed-scale convolutions to such heterogeneous contexts constrains the parameter estimator to fully exploit their complementary properties. From this perspective, the limitation lies not in insufficient context, but in the inadequate extraction and utilization of context during parameter estimation.

\section{Method}

\begin{figure}[t]
  \centering
  \includegraphics[width=0.95 \linewidth]{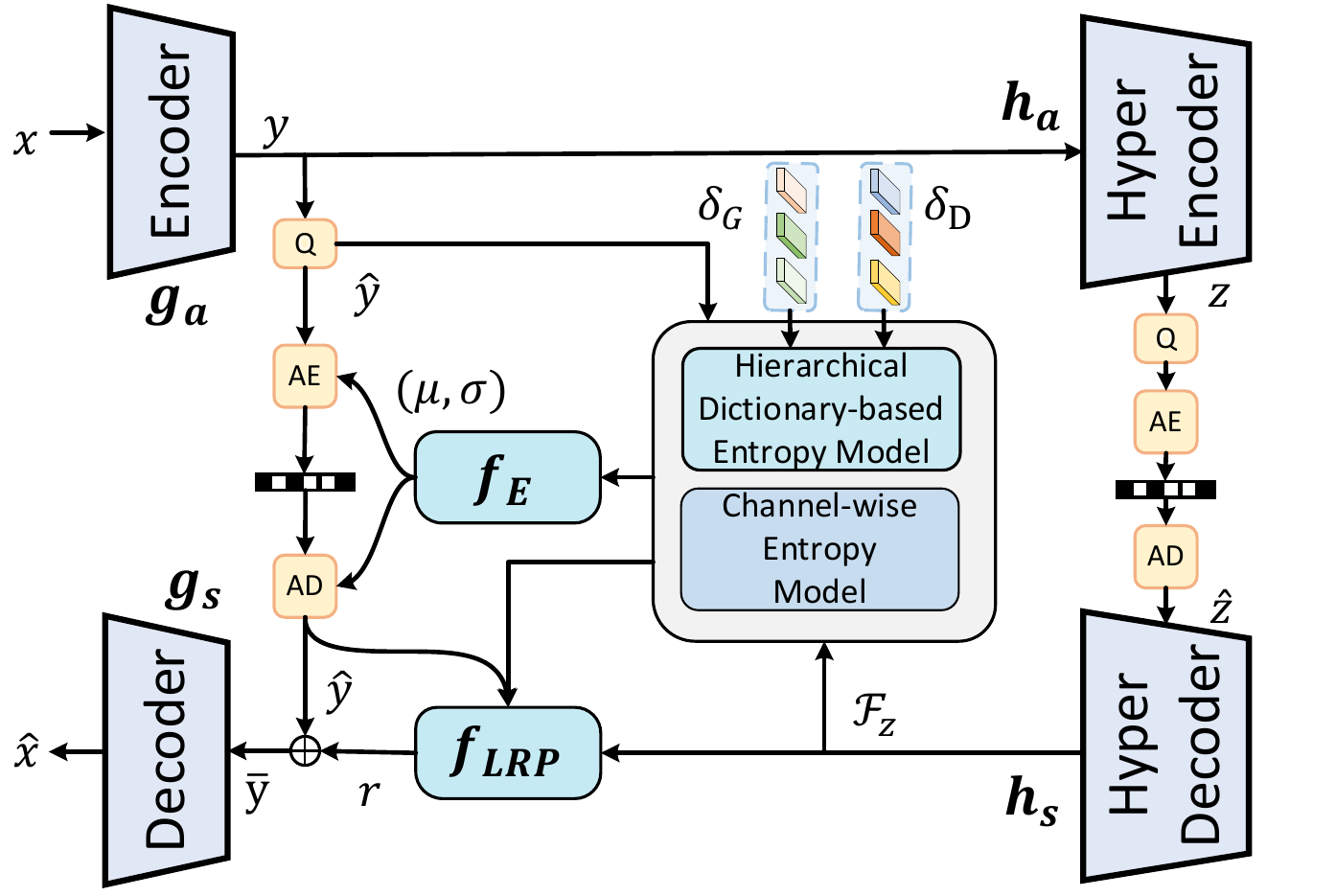}
  \caption{
  Overview of the proposed HiDE framework.
  HiDE integrates hierarchical dictionaries ($\delta_{G}, \delta_{D}$) to retrieve external priors, which are fused with the hyperprior $\mathcal{F}_z$ and channel-wise autoregressive context. 
  The aggregated context guides the entropy parameter estimator $f_E$ and the latent residual predictor $f_{LRP}$ to refine reconstruction quality and improve entropy modeling accuracy.
  }
  \label{fig:pipeline}
\end{figure}

\begin{figure*}[t]
  \centering
  \includegraphics[width=0.9\linewidth]{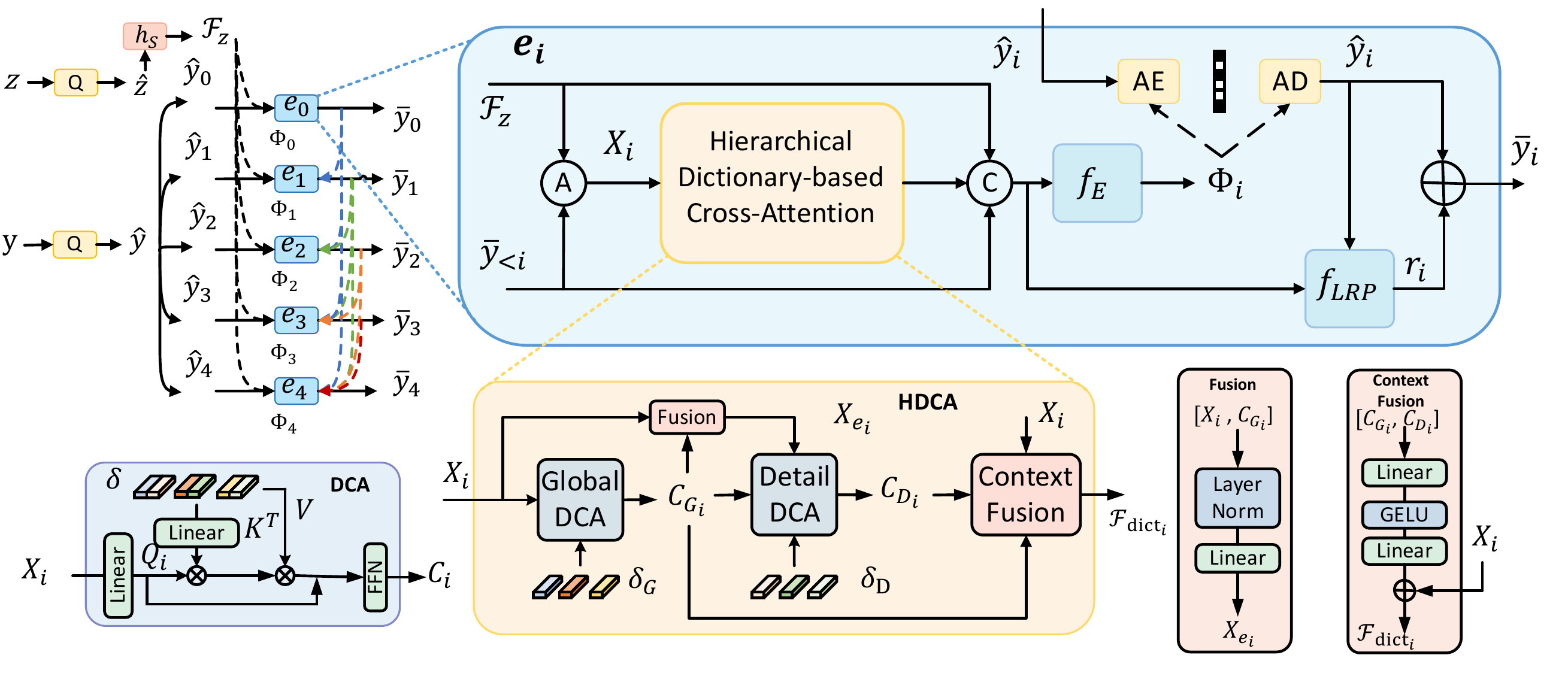}
  \caption{
  Overview of the proposed Hierarchical Dictionary-based Entropy Model, which consists of Hierarchical Dictionary-based Context Model (HD) and Context-aware Parameter Estimation (CaPE).
  Top-left: slice-wise entropy model that aggregates hyperprior $\mathcal{F}_z$ and autoregressive context $\bar{y}_{<i}$.
  Top-right (blue box $e_i$): internal structure of the $i$-th slice network employing the Hierarchical Dictionary Cross-Attention (HDCA) module to retrieve external priors for parameter estimation ($f_E$) and residual prediction ($f_{LRP}$).
  Bottom (yellow box): HDCA performs a two-stage retrieval: first querying the \textbf{Global Dictionary} $\delta_G$ for global structural priors $\mathcal{C}_{G_i}$, and then querying the \textbf{Detail Dictionary} $\delta_D$ for local detailed texture priors $\mathcal{C}_{D_i}$ conditioned on the global context.
  Bottom-right (pink box): fusion stages generating intermediate $X_{e_i}$ and final dictionary-aware context $\mathcal{F}_{dict_i}$.
  }
  \label{fig:HD}
\end{figure*}

\subsection{Preliminaries}

As illustrated in Figure \ref{fig:pipeline}, the encoder $g_a$ transforms the input image $x$ into a continuous latent representation $y = g_a(x)$. 
To support entropy coding, a hyperprior module extracts side information $z = h_a(y)$, which is quantized into $\hat{z} = Q(z)$ and encoded into the bitstream.
Here, $Q$ represents the rounding operation. 
Based on the decoded hyperprior $\hat{z}$, the hyper-decoder $h_s$ produces the hyperprior context feature $\mathcal{F}_z = h_s(\hat{z})$.
Given the predicted mean $\mu$, $y$ is quantized to $\hat{y} = Q(y-\mu) + \mu$ and $Q(y-\mu)$ is losslessly compressed.
Latent Residual Prediction (LRP) estimates the quantization error (residual) $r \approx y - \hat{y}$ and refines the latent representation as $\bar{y} = \hat{y} + r$.
Finally, the decoder $g_s$ reconstructs the image $\hat{x} = g_s(\bar{y})$.
The framework follows the rate-distortion optimization, 
\begin{equation}
  \mathcal{L} = \mathcal{R}(\hat{y}) + \mathcal{R}(\hat{z}) + \lambda \mathcal{D}(x, \hat{x}),
  \label{eq:loss}
\end{equation}
where $\mathcal{R}(\cdot)$ estimates the bitrate using the entropy model, and $\mathcal{D}(\cdot)$ measures reconstruction distortion.
Since the main contribution of HiDE lies in entropy modeling, we adopt the backbone architecture of DCAE~\citep{Lu2025} for $g_a$, $g_s$, $h_a$, and $h_s$.

The bitrate $\mathcal{R}(\hat{y})$ primarily depends on the accuracy of the conditional probability $p(\hat{y}|\cdot)$. 
Following the Gaussian assumption \citep{Balle2018, Minnen_18}, the conditional density is modeled as
$p_{\hat{y}}(\hat{y}|\hat{z}) 
= \prod_i \big(\mathcal{N}(\mu_i, \sigma_i) * \mathcal{U}(-\tfrac{1}{2}, \tfrac{1}{2})\big)(\hat{y}_i)$,
where $\mathcal{N}(\mu_i, \sigma_i)$ is a Gaussian distribution with mean $\mu_i$ and standard deviation $\sigma_i$, 
and $*$ denotes convolution with the unit uniform distribution $\mathcal{U}(-\tfrac{1}{2}, \tfrac{1}{2})$, which accounts for quantization noise.
To capture channel dependencies, $y$ is divided into $s$ channel slices $\{y_0, \dots, y_{s-1}\}$.
For each slice $i$, the context includes the hyperprior feature $\mathcal{F}_z$ and the previously decoded slices $\bar{y}_{<i}$.
The parameter estimation network $f_E$ predicts $\Phi_i = (\mu_i, \sigma_i)$, and the latent residual predictor $f_{LRP}$ estimates the quantization residual $r_i$ as:
\begin{equation}
  \begin{aligned}
    \mu_i, \sigma_i &= f_E(\mathcal{F}_z, \bar{y}_{<i}, \mathcal{F}_{dict_i}),\\
    r_i &= f_{LRP}(\mathcal{F}_z, \bar{y}_{<i}, \mathcal{F}_{dict_i}, \hat{y}_i),
  \end{aligned}
  \label{eq:entropy_parameter_estimation}
\end{equation}
where $0 \le i < s$, and $\mathcal{F}_{dict_i}$ denotes the hierarchical dictionary context introduced below.
Figure~\ref{fig:HD} presents an overview of the proposed Hierarchical Dictionary-based Entropy Model, which comprises two key components: the Hierarchical Dictionary-based Context Model (HD) and the Context-aware Parameter Estimation module (CaPE).

\subsection{Hierarchical Dictionary-based Context Modeling}

To effectively exploit external priors and alleviate dictionary representational collapse, we propose the hierarchical dictionary-based context model (HD) that decomposes external knowledge into complementary global and local components retrieved in a coarse-to-fine manner.

Two learnable dictionaries are constructed and shared between the encoder and decoder.
The global structural dictionary ${\delta}_G \in \mathbb{R}^{N_G \times C_d}$ is designed to capture global patterns and long-range dependencies, and the local detail dictionary ${\delta}_D \in \mathbb{R}^{N_D \times C_d}$ focuses on fine-grained textures and local dependencies.
$N_G, N_D$ denote the numbers of entries and $C_d$ represents the channel dimension of each dictionary entry. Both dictionaries are optimized jointly within LIC.

In the slice-wise context model, the latent representation $y$ is partitioned into channel-wise slices $y_i$.
For the $i$-th slice, the input context $X_i$ is formed by aggregating the hyperprior feature $\mathcal{F}_z$ and the previously decoded slices $\bar{y}_{<i}$.
The aggregation followed with \citep{Lu2025}.

The hierarchical retrieval is performed in two sequential stages that progressively refine the external prior.
In the first stage, the global dictionary is queried using $Q_{G_i}$ to retrieve the global context feature $\mathcal{C}_{G_i}$ via cross-attention.
The global dictionary serves as both keys and values, providing coarse structural references.
We employ a multi-head cross-attention mechanism,
\begin{equation}
  Q_{G_i} = X_i W_Q^{G}, \quad
  K_{G} = \boldsymbol{\delta}_G W_K^{G}, \quad
  V_{G} = \boldsymbol{\delta}_G,
  \end{equation}
\begin{equation}
    \mathcal{C}_{G_i} = \text{Softmax}\left(\frac{Q_{G_i} K_G^T}{\tau_i}\right) V_G,
\end{equation}
where $W_Q^{G}$ and $W_K^{G}$ are learnable projection matrices.
The $\tau_i$ is a learnable temperature parameter that controls the sharpness of the attention distribution.
By providing a structural prior, this stage reduces uncertainty in subsequent retrieval within a coherent framework.

The second stage retrieves detail texture conditioned on the global context.
As shown in the fusion block (Figure \ref{fig:HD}, bottom-right), the enhanced query $X_{e_i}$ is constructed by fusing the original context $X_i$ with the global prior $\mathcal{C}_{G_i}$ via a linear projection followed by LayerNorm operator:

\begin{equation}
  X_{e_i} = \text{LayerNorm}([X_i, \mathcal{C}_{G_i}]W_{proj}).
  \label{eq:enhanced_query}
\end{equation}

Conditioning detail retrieval on the global context constrains texture selection to be structurally consistent, leading to more stable and discriminative dictionary utilization.
The detail attention proceeds as:
\begin{equation}
    Q_{D_i} = X_{e_i} W_Q^D, \quad K_D = {\delta}_D W_K^D, \quad V_D = {\delta}_D, 
\end{equation}

\begin{equation}
    \mathcal{C}_{D_i} = \text{Softmax}\left(\frac{Q_{D_i} K_D^T}{\tau_i}\right) V_D.
\end{equation}

Finally, the retrieved global and detail contexts are integrated to form the dictionary-aware representation. 
As shown in Figure \ref{fig:HD} (bottom-right), we first employ a lightweight linear layer to fuse the heterogeneous dictionary features $[\mathcal{C}_{G_i}, \mathcal{C}_{D_i}]$.
To ensure that these external priors serve as a refinement without losing the original internal context, we mathematically formulate this fusion with a residual connection from the input $X_i$:
\begin{equation}
\mathcal{F}_{\text{dict}_i} = \phi([\mathcal{C}_{G_i}, \mathcal{C}_{D_i}] W_1) W_2 + X_i,
\label{eq:hdict_fusion}
\end{equation}
where $W_1, W_2$ are projection matrices and $\phi$ is the GELU activation.
This residual design allows the gradients to propagate effectively and ensures the model explicitly learns to enrich the internal context with external knowledge.

\subsection{Context-aware Parameter Estimation}

\begin{figure}[t]
  \centering
  \includegraphics[width=\linewidth]{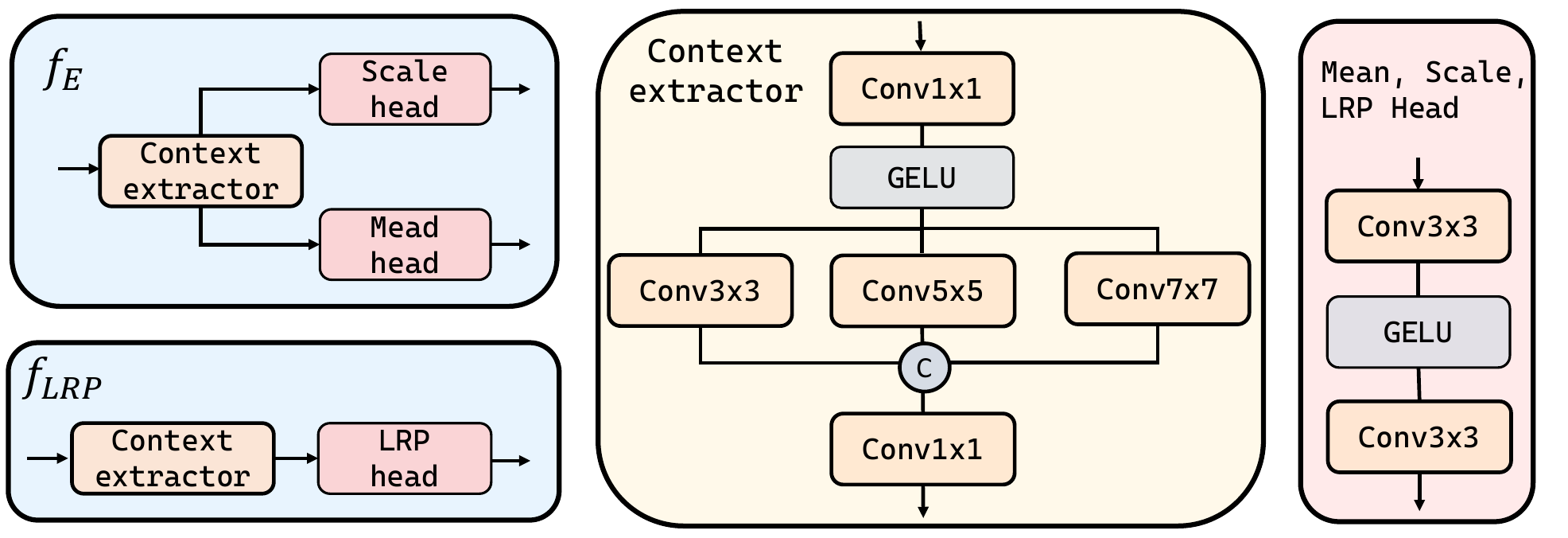}
  \caption{
  Illustration of the proposed Context-aware Parameter Estimation (CaPE) module.
  Left: CaPE was employed in two module in entropy model. 
  Firstly, CaPE served as the parameter estimator $f_E$ extracts a shared context representation to predict both $\mu$ and $\sigma$.
  Then $f_{LRP}$ predicts the quantization residual $r$.
  Middle: The internal structure of the context extractor. 
  Right: The task-specific heads for mean, scale, and residual prediction, implemented with lightweight stacked $3\times3$ convolutions and GELU activations.
  }
  \label{fig:mrfpe}
\end{figure}

To effectively exploit and interpret diverse priors from the hyperprior, channel-wise autoregressive context, and dictionary contexts, we introduce the context-aware parameter estimation (CaPE) module as illustrated in Figure~\ref{fig:mrfpe}.  
CaPE enhances conventional parameter estimators by employing a parallel branches with multi-receptive field design that dynamically captures correlations across heterogeneous context.

Given the aggregated feature $\mathcal{S} = [\mathcal{F}_z, \bar{y}_{<i}, \mathcal{F}_{dict_i}]$, CaPE first applies a $1{\times}1$ convolution to project $\mathcal{S}$ into a lower-dimensional feature space $\mathcal{S}_{proj}$, followed by three parallel convolutional branches with kernel sizes $k \in \{3, 5, 7\}$:
\begin{equation}
  F_k = \phi(\text{Conv}_{k \times k}(\mathcal{S}_{proj})), \quad k \in \{3, 5, 7\},
\end{equation}
where $\phi$ denotes the GELU activation.  
The outputs of these branches capture both local and global dependencies, and are concatenated and fused via another $1{\times}1$ convolution:
\begin{equation}
  \mathcal{F}_{ctx} = \text{Conv}_{1\times1}(\text{Concat}([F_3, F_5, F_7])).
\end{equation}

The fused representation $\mathcal{F}_{ctx}$ serves as the shared context for parameter estimation.  
Specifically, $f_E$ predicts the Gaussian distribution parameters $(\mu, \sigma)$ through two lightweight task-specific heads $\mathcal{H}_{\mu}$ and $\mathcal{H}_{\sigma}$.
For the latent residual prediction $f_{LRP}$ employs another context extractor with the LRP head $\mathcal{H}_{lrp}$ to estimate the quantization residual $r$:
\begin{equation}
  \mu = \mathcal{H}_{\mu}(\mathcal{F}_{ctx}), \quad 
  \sigma = \mathcal{H}_{\sigma}(\mathcal{F}_{ctx}), \quad
  r = \mathcal{H}_{lrp}(\mathcal{F}_{ctx}).
\end{equation}

Benefit from parallel branched multi-receptive fields based context extractor and task-specific predict heads, CaPE enables more accurate entropy parameter prediction and residual correction, substantially improving compression performance when combined with the hierarchical dictionary priors.

\section{Experiments}

\subsection{Experimental Settings}

Our model is trained on 300k images sampled from the OpenImage dataset~\citep{krasin2017openimages}.  
During training, we randomly crop $256 \times 256$ patches and use a batch size of 16.  
The optimization is performed using the Adam optimizer with an initial learning rate of $1\times10^{-4}$ for 80 epochs, which is subsequently decayed to $1\times10^{-5}$ for another 20 epochs.  
To obtain different rate–distortion (RD) trade-offs, the weighting parameter $\lambda$ in Eq.~(\ref{eq:loss}) is varied among $\{0.0018, 0.0035, 0.0067, 0.013, 0.025, 0.05\}$.  
Training is performed on two NVIDIA RTX~4090 GPUs, taking approximately 16~days per bitrate.
Six rate points are obtained by training independent models for $\lambda \in \{0.0035, 0.05\}$ and fine-tuning additional models for 5~epochs initialized from the pre-trained checkpoints (at the 95th epoch).
Additional implementation configurations and ablation results are provided in the supplementary material.

\subsection{Comparisons with State-of-the-Art Methods}

\begin{figure*}
  \centering
  \begin{subfigure}[b]{0.33\linewidth}
      \centering
      \includegraphics[width=\linewidth]{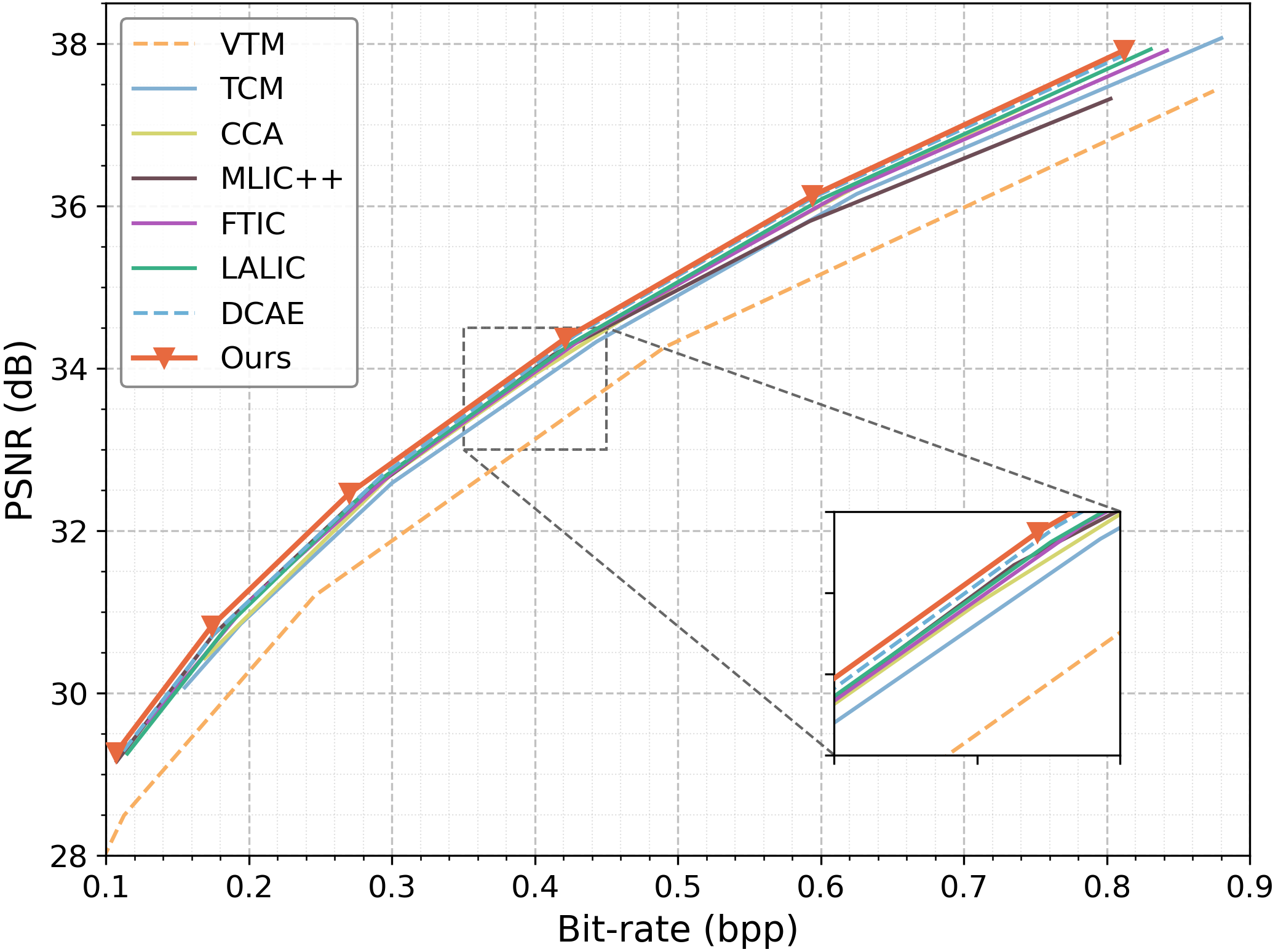} 
      \label{fig:img_a}
  \end{subfigure}
  \hfill
  \begin{subfigure}[b]{0.33\linewidth}
      \centering
      \includegraphics[width=\linewidth]{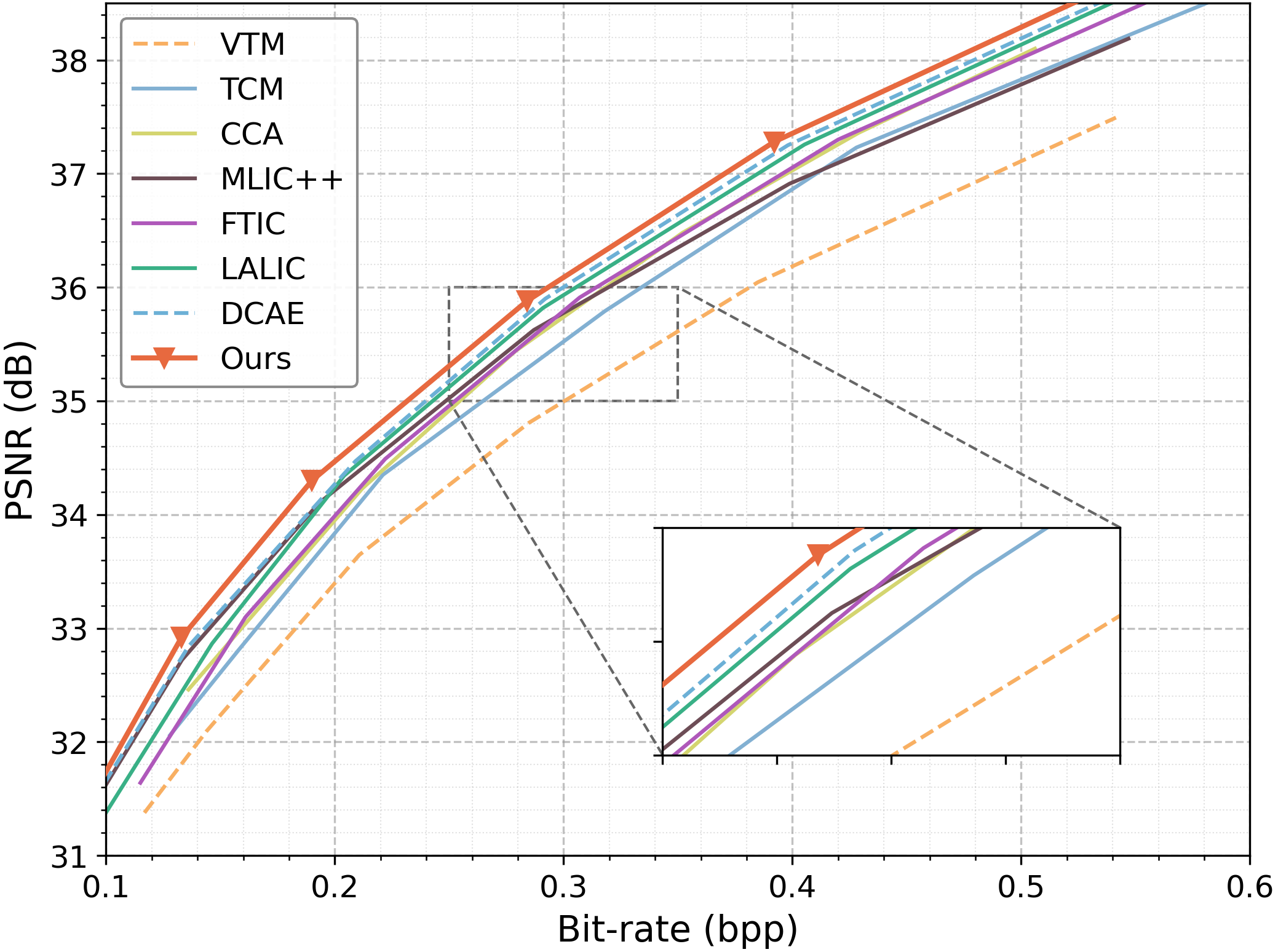} 
      \label{fig:img_b}
  \end{subfigure}
  \hfill
  \begin{subfigure}[b]{0.33\linewidth}
      \centering
      \includegraphics[width=\linewidth]{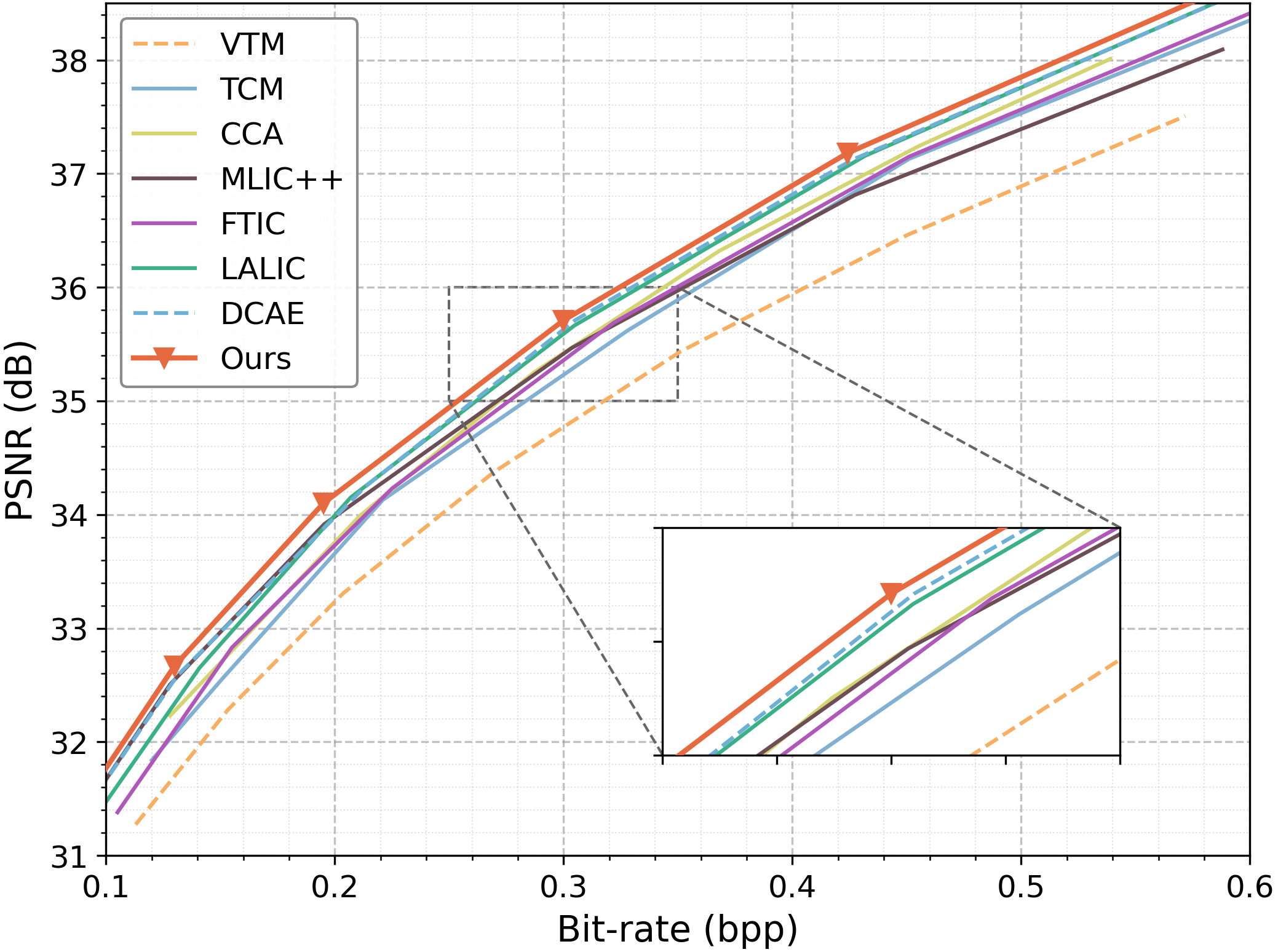} 
      \label{fig:img_c}
  \end{subfigure}
  \caption{Rate-distortion curve on three benchmark datasets: Kodak, Tecnick, and CLIC Professional Validation (From left to right).}
  \label{fig:rd_comparison}
\end{figure*}

\begin{table*}
  \centering
  \begin{threeparttable}
  \small
  \begin{tabular}{l|cccc ccc c c}
  \toprule
  \multirow{3}{*}{Model} & \multicolumn{4}{c}{BD-Rate (\%) w.r.t. VTM-12.0} & \multicolumn{3}{c}{Latency (ms)} & \multirow{3}{*}{Params (M)} & \multirow{3}{*}{GFLOPs} \\
  \cmidrule(lr){2-5} \cmidrule(lr){6-8}
   & \multicolumn{2}{c}{Kodak} & Tecnick & CLIC & \multirow{2}{*}{Total} & \multirow{2}{*}{Enc.} & \multirow{2}{*}{Dec.} & & \\
  \cmidrule(lr){2-3}
   & PSNR & MS-SSIM & PSNR & PSNR & & & & & \\
  \midrule
  TCM (CVPR'23) & -11.97 & -- & -11.95 & -11.96 & 220 & 120 & 100 & 75.9 & 700.65 \\
  MLIC+ (ACMMM'23) & -13.19 & -- & -17.47 & -16.45 & -- & -- & -- & -- & -- \\
  MLIC++ (ICMLW'23) & -15.09 & -- & -18.68 & -16.84 & 235.4 & 106.5 & 128.9 & 116.7 & 615.93 \\
  FTIC (ICLR'24) & -14.84 & -54.30 & -15.24 & -13.58 & 6846.6 & 1727.7 & 3391.9 & 69.78 & 245.46 \\
  CCA (NeurIPS'24) & -13.84 & -- & -15.34 & -14.67 & 110 & 72 & 38 & 64.9 & 615.93 \\
  LALIC (CVPR'25) & -15.49 & -54.00 & -18.50 & -18.08 & 210 & 143.9 & 66.1 & 66.13 & 303.18 \\
  DCAE (CVPR'25) & -16.83 & -55.66 & -21.28 & -19.59 & 128 & 63 & 65 & 119.4 & 426.92 \\
  \textbf{HiDE (Ours)} & \textbf{-18.50} & \textbf{-56.49} & \textbf{-24.01} & \textbf{-21.99} & 134 & 66 & 68 & 134.9 & 447.64 \\
  \bottomrule
  \end{tabular}
  \end{threeparttable}
  \caption{Comparison with state-of-the-art compression methods.  
  BD-Rate is reported with respect to VTM-12.0.  
  Latency, parameter count, and GFLOPs are measured on the Kodak dataset using one NVIDIA RTX 4090 GPU.}
  \label{tab:sota_comparison}
\end{table*}

We evaluate the proposed HiDE framework on three widely used benchmarks: Kodak~\citep{kodak}, Tecnick~\citep{asuni2014testimages}, and the CLIC professional validation dataset~\citep{CLIC}.  
The comparison includes the conventional codec VVC (VTM-12.1)~\citep{dominguez2022versatile} and recent state-of-the-art learned image compression (LIC) models, namely TCM~\citep{Liu23}, MLIC+~\citep{Jiang23}, MLIC++~\citep{jiang2023mlic}, CCA~\citep{Li_ftic_24}, FTIC~\citep{Han24}, LALIC~\citep{Lalic_25}, and DCAE~\citep{Lu2025}.  

Figure~\ref{fig:rd_comparison} shows the rate–distortion (RD) curves across all datasets, and Table~\ref{tab:sota_comparison} reports the corresponding BD-Rate~\citep{bjontegaard2001calculation} savings and model complexity metrics.  
Overall, HiDE consistently achieves the lowest BD-Rate on all three benchmarks, surpassing DCAE and other competitors by a notable margin.  
The performance advantage is particularly pronounced on high-resolution datasets such as Tecnick (1K) and CLIC (2K), highlighting the benefit of hierarchical prior modeling in capturing both global structures and fine textures.  
In terms of computational efficiency, HiDE achieves these gains with only marginal increases in parameters and GFLOPs, while maintaining comparable latency.

\subsection{Ablation Studies}

To examine the contribution of each proposed component, we conduct ablation studies focusing on the hierarchical dictionary-based cross-attention (HD) and the context-aware parameter estimation (CaPE) modules.  
All ablation models are built upon DCAE~\citep{Lu2025} and trained on a reduced dataset of 14k images from VOC~\citep{everingham2010pascal} and DIV2K~\citep{Agustsson_2017_CVPR_Workshops}.  
Each model is trained for 300 epochs with a batch size of 8, and evaluated on the Kodak dataset for fair comparison.

\begin{table}[t]
  \centering
  \begin{tabular}{lcc}
  \toprule
  Model & BD-Rate (\%) & Params (M) \\
  \midrule
  +HD   & -1.35 & 139.4 \\
  +CaPE & -2.82 & 114.4 \\
  HD + CaPE (HiDE) & -3.81 & 134.9 \\
  \bottomrule
  \end{tabular}
  \caption{Ablation of hierarchical dictionary (HD) and context-aware parameter estimation (CaPE) on the Kodak dataset. DCAE~\citep{Lu2025} is used as the baseline.}
  \label{tab:ablation1}
\end{table}

\begin{figure*}
  \centering
  \includegraphics[width=\linewidth]{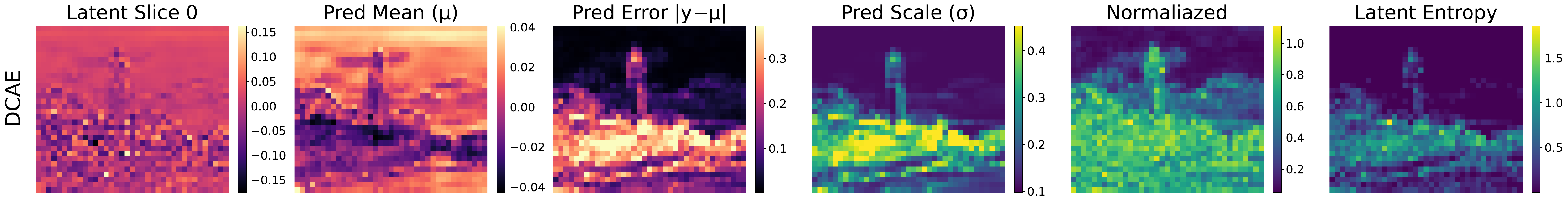}
  \includegraphics[width=\linewidth]{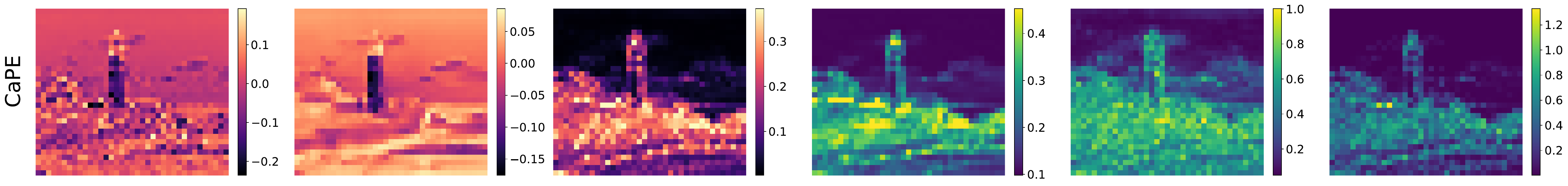}
  \includegraphics[width=\linewidth]{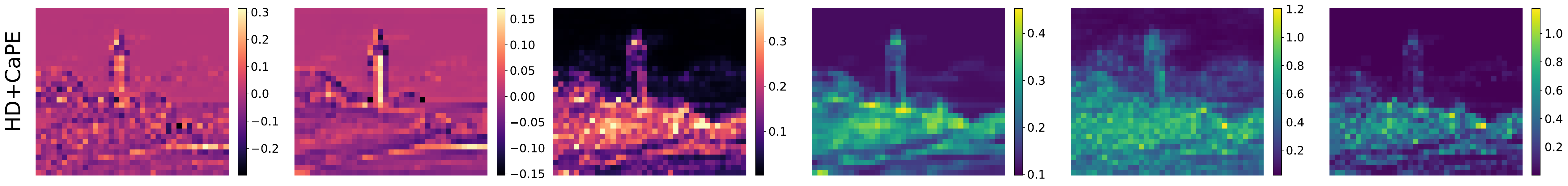}
  \caption{Visualization of latent representations and predicted distribution parameters.  
  From top to bottom: baseline DCAE, CaPE-only variant, and full HiDE (HD+CaPE).  
  From left to right: latent slice, predicted mean, absolute prediction error $(y-\mu)$, predicted scale $\sigma$, normalized residual $(y-\mu)/\sigma$, and latent entropy map.}
  \label{fig:parameter_Visualizations}
\end{figure*}

\subsubsection{Effectiveness of the Proposed Components}

Table~\ref{tab:ablation1} quantifies the impact of each module.  
Replacing the single-level dictionary in DCAE with our hierarchical design (+HD) achieves a 1.35\% BD-rate reduction over DCAE, demonstrating that decomposing external priors into global and detail dictionaries mitigates representational competition.  
Substituting the standard fixed-scale estimator with the proposed CaPE module (+CaPE) further improves compression efficiency by 2.82\%, while also reducing the parameter count from 119.4M to 114.4M.  
As visualized in Figure~\ref{fig:entry_usage}, our hierarchical dictionaries exhibit more balanced utilization compared to the flat structure of DCAE, confirming improved representational diversity.  
When both components are combined, HiDE achieves the largest overall gain of 3.81\%, validating their strong complementarity.

\begin{table}[t]
  \centering
  \begin{threeparttable}
  \begin{tabular}{lcc}
  \toprule
  \textbf{G}-\textbf{D} & BD-Rate (\%) & BD-PSNR (dB) \\
  \midrule
  32–96 & -1.26 & 0.060 \\
  64–64 & \textbf{-1.35} & \textbf{0.065} \\
  \bottomrule
  \end{tabular}
  \end{threeparttable}
  \caption{Impact of different global–detail (\textbf{G–D}) dictionary size configurations under a fixed total dictionary capacity. DCAE~\citep{Lu2025} serves as the baseline.}
  \label{tab:dictionary_size}
\end{table}

\subsubsection{Effect of Dictionary Size}

To analyze the impact of capacity allocation between the global and detail dictionaries, we vary their sizes while maintaining the same total number of entries.  
As shown in Table~\ref{tab:dictionary_size}, both configurations outperform the baseline DCAE, achieving BD-rate reductions of 1.26\% and 1.35\%.  
The similar performance across configurations suggests that the improvement mainly stems from the hierarchical decomposition rather than the specific dictionary size ratio.  
Accordingly, we adopt the balanced configuration (64–64) as the default setting for HiDE.

\subsubsection{Generalization of Parameter Estimation}

\begin{table}
  \centering
  \begin{tabular}{lccc}
  \toprule
  Model & BD-Rate (\%) & Params (M) & Latency (ms) \\
  \midrule
  TCM & -5.3 & 45.2 & 93 \\
  TCM + CaPE & -5.6 & 55.0 & 94 \\
  \bottomrule
  \end{tabular}
  \caption{Generalization of the context-aware parameter estimation (CaPE) module when applied to the small version of TCM~\citep{Liu23} framework. Results are reported with \citep{Cheng20} as the anchor.}
  \label{tab:Generalization}
\end{table}

To evaluate the generality of CaPE, we integrate it into the small version of TCM~\citep{Liu23} model by replacing its original parameter estimation module.  
As reported in Table~\ref{tab:Generalization}, CaPE yields an additional BD-rate reduction of 0.3\% with competitive latency overhead, indicating that its benefit extends beyond dictionary-based architectures.  
Although the improvement in TCM is smaller compared to DCAE, this is expected since TCM relies solely on channel-sliced internal contexts, while dictionary-based models like DCAE benefit more from context-aware estimation due to their richer contextual interactions.

\subsection{Analysis of Parameter Estimation in LIC}

Figure~\ref{fig:parameter_Visualizations} visualizes the distribution parameters predicted by DCAE, the CaPE-only variant, and HiDE (HD+CaPE).  
We display the latent slice with the highest entropy from the Kodak image \emph{kodim21}.  
While the latent maps $y$ and predicted means $\mu$ (first and second columns) appear visually similar across models, the prediction error $(y-\mu)$ (third column) reveals a clear reduction in residual magnitude for HiDE.  
This improvement is accompanied by smaller predicted scales $\sigma$ (fourth column), reflecting lower uncertainty and more confident estimation.  
We further assess the modeling capacity via the normalized residuals $\frac{(y - \mu)}{\sigma}$ in the fifth column.
For the baseline DCAE, stronger structural correlations persist in the normalized domain with the outline of the lighthouse clearly visible. 
In contrast, our method substantially mitigates these structural dependencies, which indicates enhanced spatial decorrelation.
Finally, the last column visualizes the spatial allocation of bitrate. HiDE produces the most compact representations, underscoring the critical role of accurate parameter estimation in optimizing coding efficiency.

\section{Conclusion}

This paper presents \textbf{HiDE}, a hierarchical dictionary-based entropy modeling framework that effectively exploits external priors for learned image compression. 
HiDE organizes external priors into global and detail dictionaries to model coarse structural patterns and fine-grained textures while alleviating representational conflicts. 
The cascaded retrieval mechanism with global conditioning ensures semantic consistency and promotes balanced utilization of external priors. 
In addition, a context-aware parameter estimation network is introduced to overcome the limitations of single-scale convolutional estimators, improving the accuracy of conditional distribution prediction. Experimental results demonstrate that HiDE consistently outperforms existing state-of-the-art methods on various benchmark datasets, validating the effectiveness of hierarchical external prior modeling for efficient entropy estimation.

\bibliographystyle{named} 
\bibliography{reference} 

@inproceedings{vqvae17,
  author       = {Aaron Van Den Oord and
                  Oriol Vinyals and
                  Koray Kavukcuoglu},
  
  title        = {Neural Discrete Representation Learning},
  booktitle    = {Proceedings of the Advances in Neural Information Processing Systems 30: Annual Conference
                  on Neural Information Processing Systems,
                  Long Beach, CA, {USA}},
  pages        = {6306--6315},
  year         = {2017},
  }

@inproceedings{vqgan21,
  author       = {Patrick Esser and
                  Robin Rombach and
                  Bj{\"{o}}rn Ommer},
  title        = {Taming Transformers for High-Resolution Image Synthesis},
  booktitle    = {Proceedings of the {IEEE} Conference on Computer Vision and Pattern Recognition, 
                  virtual},
  pages        = {12873--12883},
  publisher    = {Computer Vision Foundation / {IEEE}},
  year         = {2021},
  month = {June},
 }

@inproceedings{Collapse25,
  title={Addressing representation collapse in vector quantized models with one linear layer},
  author={Zhu, Yongxin and Li, Bocheng and Xin, Yifei and Xia, Zhihua and Xu, Linli},
  booktitle={Proceedings of the IEEE/CVF International Conference on Computer Vision},
  pages={22968--22977},
  year={2025}
}

@inproceedings{Balle17,
  author       = {Johannes Ball{\'{e}} and
                  Valero Laparra and
                  Eero P. Simoncelli},
  title        = {End-to-end Optimized Image Compression},
  booktitle    = {Proceedings of the 5th International Conference on Learning Representations, 
                  Toulon, France,},

  year         = {2017},
  month = {April},
 }

@inproceedings{Balle2018,
  author       = {Johannes Ball\'{e} and
                  David Minnen and
                  Saurabh Singh and
                  Sung Jin Hwang and
                  Nick Johnston},
  title        = {Variational image compression with a scale hyperprior},
  booktitle    = {Proceedings of the 6th International Conference on Learning Representations, 
                  Vancouver, BC, Canada},

  year         = {2018},
  month = {April},
  }

@article{Balle21,
  author       = {Johannes Ball{\'{e}} and
                  Philip A. Chou and
                  David Minnen and
                  Saurabh Singh and
                  Nick Johnston and
                  Eirikur Agustsson and
                  Sung Jin Hwang and
                  George Toderici},
  title        = {Nonlinear Transform Coding},
  journal      = {{IEEE} J. Sel. Top. Signal Process.},
  volume       = {15},
  number       = {2},
  pages        = {339--353},
  year         = {2021},
 }

@inproceedings{MinnenTSHC18,
  author       = {David Minnen and
                  George Toderici and
                  Saurabh Singh and
                  Sung Jin Hwang and
                  Michele Covell},
  title        = {Image-Dependent Local Entropy Models for Learned Image Compression},
  booktitle    = {Proceedings of the 2018 {IEEE} International Conference on Image Processing, 
                  Athens, Greece,},
  pages        = {430--434},
  publisher    = {{IEEE}},
  year         = {2018},
  month = {October},
  }

@inproceedings{Minnen_18,
  author       = {David Minnen and
                  Johannes Ball{\'{e}} and
                  George Toderici},
  
  title        = {Joint Autoregressive and Hierarchical Priors for Learned Image Compression},
  booktitle    = {Proceedings of the Advances in Neural Information Processing Systems 31: Annual Conference
                  on Neural Information Processing Systems 2018, NeurIPS 2018, December
                  3-8, 2018, Montr{\'{e}}al, Canada},
  pages        = {10794--10803},
  year         = {2018},
  url          = {https://proceedings.neurips.cc/paper/2018/hash/53edebc543333dfbf7c5933af792c9c4-Abstract.html},
  bibsource    = {dblp computer science bibliography, https://dblp.org}
}

@inproceedings{MinnenS20,
  author       = {David Minnen and
                  Saurabh Singh},
  title        = {Channel-Wise Autoregressive Entropy Models for Learned Image Compression},
  booktitle    = {Proceedings of the {IEEE} International Conference on Image Processing, 
                  Abu Dhabi, United Arab Emirates, },
  pages        = {3339--3343},
  publisher    = {{IEEE}},
  year         = {2020},
  month = {October},
  }

@inproceedings{Cheng20,
  author       = {Zhengxue Cheng and
                  Heming Sun and
                  Masaru Takeuchi and
                  Jiro Katto},
  title        = {Learned Image Compression With Discretized Gaussian Mixture Likelihoods
                  and Attention Modules},
  booktitle    = {Proceedings of the 2020 {IEEE/CVF} Conference on Computer Vision and Pattern Recognition,
                  Seattle, WA, USA},
  pages        = {7936--7945},
  publisher    = {Computer Vision Foundation / {IEEE}},
  year         = {2020},
  month = {June},
  }

@inproceedings{He_21,
  author       = {Dailan He and
                  Yaoyan Zheng and
                  Baocheng Sun and
                  Yan Wang and
                  Hongwei Qin},
  title        = {Checkerboard Context Model for Efficient Learned Image Compression},
  booktitle    = {Proceedings of the {IEEE} Conference on Computer Vision and Pattern Recognition, 
                  virtual,},
  pages        = {14771--14780},
  publisher    = {Computer Vision Foundation / {IEEE}},
  year         = {2021},
  month = {June},
  }

@inproceedings{He_22,
  author       = {Dailan He and
                  Ziming Yang and
                  Weikun Peng and
                  Rui Ma and
                  Hongwei Qin and
                  Yan Wang},
  title        = {{ELIC:} Efficient Learned Image Compression with Unevenly Grouped
                  Space-Channel Contextual Adaptive Coding},
  booktitle    = {Proceedings of the {IEEE/CVF} Conference on Computer Vision and Pattern Recognition,
                   New Orleans, LA, USA, },
  pages        = {5708--5717},
  publisher    = {Computer Vision Foundation / {IEEE}},

  year         = {2022},
  month = {June},
  
 }

@inproceedings{Liu23,
  author       = {Jinming Liu and
                  Heming Sun and
                  Jiro Katto},
  title        = {Learned Image Compression with Mixed Transformer-CNN Architectures},
  booktitle    = {Proceedings of the {IEEE/CVF} Conference on Computer Vision and Pattern Recognition,
                  Vancouver, BC, Canada,},
  pages        = {14388--14397},
  publisher    = {Computer Vision Foundation / {IEEE}},
  year         = {2023},
  month = {June},
  }

@inproceedings{Jiang23,
  author       = {Wei Jiang and
                  Jiayu Yang and
                  Yongqi Zhai and
                  Peirong Ning and
                  Feng Gao and
                  Ronggang Wang},
  title        = {{MLIC:} Multi-Reference Entropy Model for Learned Image Compression},
  booktitle    = {Proceedings of the 31st {ACM} International Conference on Multimedia,
                  Ottawa, ON, Canada, },
  pages        = {7618--7627},
  publisher    = {{ACM}},
  year         = {2023},
  
 }

@article{jiang2023mlic,
  author       = {Wei Jiang and
                  Ronggang Wang},
  title        = {{MLIC++:} Linear Complexity Multi-Reference Entropy Modeling for Learned
                  Image Compression},
  journal      = {ICML 2023 Workshop Neural Compression: From Information Theory to Applications},
  
  year         = {2023},
  url          = {https://openreview.net/forum?id=hxIpcSoz2t},
  
}

@inproceedings{Li_ftic_24,
  author       = {Han Li and
                  Shaohui Li and
                  Wenrui Dai and
                  Chenglin Li and
                  Junni Zou and
                  Hongkai Xiong},
  title        = {Frequency-Aware Transformer for Learned Image Compression},
  booktitle    = {Proceedings of the 12th International Conference on Learning Representations,
                  Vienna, Austria},
  year         = {2024},
  month = {May},
 }

@inproceedings{Han24,
  author       = {Minghao Han and
                  Shiyin Jiang and
                  Shengxi Li and
                  Xin Deng and
                  Mai Xu and
                  Ce Zhu and
                  Shuhang Gu},
  title        = {Causal Context Adjustment Loss for Learned Image Compression},
  booktitle    = {Proceedings of the Advances in Neural Information Processing Systems 38: Annual Conference
                  on Neural Information Processing Systems 2024, 
                  Vancouver, BC, Canada,},
  year         = {2024},
  month = {December},
 }

@inproceedings{Lalic_25,
  author       = {Donghui Feng and
                  Zhengxue Cheng and
                  Shen Wang and
                  Ronghua Wu and
                  Hongwei Hu and
                  Guo Lu and
                  Li Song},
  title        = {Linear Attention Modeling for Learned Image Compression},
  booktitle    = {Proceedings of the {IEEE/CVF} Conference on Computer Vision and Pattern Recognition,
                  Nashville, TN, USA},
  pages        = {7623--7632},
  publisher    = {Computer Vision Foundation / {IEEE}},
  year         = {2025},
  month = {June},
}

@inproceedings{zeng2025mambaic,
  title={MambaIC: State Space Models for High-Performance Learned Image Compression},
  author={Zeng, Fanhu and Tang, Hao and Shao, Yihua and Chen, Siyu and Shao, Ling and Wang, Yan},
  booktitle={Proceedings of the {IEEE/CVF} Conference on Computer Vision and Pattern Recognition,
             Nashville, TN, USA},
  pages={18041--18050},
  year={2025},
  month = {June},
}

@inproceedings{Zhong20,
  author       = {Zhisheng Zhong and
                  Hiroaki Akutsu and
                  Kiyoharu Aizawa},
  title        = {Channel-Level Variable Quantization Network for Deep Image Compression},
  booktitle    = {Proceedings of the Twenty-Ninth International Joint Conference on
                  Artificial Intelligence},
  pages        = {467--473},
  publisher    = {ijcai.org},
  year         = {2020},
 }

@inproceedings{Zhu22,
  author       = {Yinhao Zhu and
                  Yang Yang and
                  Taco Cohen},
  title        = {Transformer-based Transform Coding},
  booktitle    = {Proceedings of the 10th International Conference on Learning Representations, 
                  Virtual Event},

  year         = {2022},
  month = {April},
}

@article{JPEG,
  author       = {Gregory K. Wallace},
  title        = {The {JPEG} Still Picture Compression Standard},
  journal      = {Commun. {ACM}},
  volume       = {34},
  number       = {4},
  pages        = {30--44},
  year         = {1991},
  }

@article{VVC2021,
  author       = {Benjamin Bross and
                  Ye{-}Kui Wang and
                  Yan Ye and
                  Shan Liu and
                  Jianle Chen and
                  Gary J. Sullivan and
                  Jens{-}Rainer Ohm},
  title        = {Overview of the Versatile Video Coding {(VVC)} Standard and its Applications},
  journal      = {{IEEE} Trans. Circuits Syst. Video Technol.},
  volume       = {31},
  number       = {10},
  pages        = {3736--3764},
  year         = {2021},
 }

@inproceedings{Lu2025,
  author       = {Jingbo Lu and
                  Leheng Zhang and
                  Xingyu Zhou and
                  Mu Li and
                  Wen Li and
                  Shuhang Gu},
  title        = {Learned Image Compression with Dictionary-based Entropy Model},
  booktitle    = {Proceedings of the {IEEE/CVF} Conference on Computer Vision and Pattern Recognition,
                  Nashville, TN, USA},
  pages        = {12850--12859},
  publisher    = {Computer Vision Foundation / {IEEE}},
  year         = {2025},
  month = {June},
  }

@inproceedings{WuCLH25,
  author       = {Siqi Wu and
                  Yinda Chen and
                  Dong Liu and
                  Zhihai He},
  title        = {Conditional Latent Coding with Learnable Synthesized Reference for
                  Deep Image Compression},
  booktitle    = {Proceedings of the AAAI-25, Sponsored by the Association for the Advancement of Artificial
                  Intelligence, Philadelphia, PA, {USA}},
  pages        = {12863--12871},
  publisher    = {{AAAI} Press},
  year         = {2025},

  }

@InProceedings{Kuang_2025_ICCV,
    author    = {Kuang, Haowei and Yang, Wenhan and Guo, Zongming and Liu, Jiaying},
    title     = {Cross-Granularity Online Optimization with Masked Compensated Information for Learned Image Compression},
    booktitle = {Proceedings of the IEEE/CVF International Conference on Computer Vision},
    month     = {October},
    year      = {2025},
    pages     = {16514-16523}
}

@article{krasin2017openimages,
  title={Openimages: A public dataset for large-scale multi-label and multi-class image classification},
  author={Krasin, Ivan and Duerig, Tom and Alldrin, Neil and Ferrari, Vittorio and Abu-El-Haija, Sami and Kuznetsova, Alina and Rom, Hassan and Uijlings, Jasper and Popov, Stefan and Veit, Andreas and others},
  journal={Dataset available from https://github. com/openimages},
  volume={2},
  number={3},
  pages={18},
  year={2017}
}

@article{Kodak,
  title={kodak lossless true color image suite},
  author={Eastman kodak},
  journal={Dataset available from https://r0k.us/graphics/kodak/},
  year={1993}
}

@article{asuni2014testimages,
  title={TESTIMAGES: a Large-scale Archive for Testing Visual Devices and Basic Image Processing Algorithms.},
  author={Asuni, Nicola and Giachetti, Andrea and others},
  booktitle={STAG},
  pages={63--70},
  year={2014}
}

@misc{CLIC,
  author={CLIC},
  title={Workshop and challenge on learned image compression and multi-class image classification.},
  journal={Dataset available from https://compression.cc/},
  year={2021}
}

@InProceedings{Agustsson_2017_CVPR_Workshops,
	author = {Agustsson, Eirikur and Timofte, Radu},
	title = {NTIRE 2017 Challenge on Single Image Super-Resolution: Dataset and Study},
	booktitle = {The IEEE Conference on Computer Vision and Pattern Recognition (CVPR) Workshops},
	month = {July},
	year = {2017}
}

@book{dominguez2022versatile,
  title={Versatile Video Coding},
  author={Dominguez, Humberto Ochoa and Rao, Kamisetty Ramamohan},
  year={2022},
  publisher={River publishers}
}

@article{bjontegaard2001calculation,
  title={Calculation of average PSNR differences between RD-curves},
  author={Bjontegaard, Gisle},
  journal={ITU-T SG16, Doc. VCEG-M33},
  year={2001}
}

@misc{everingham2010pascal,
      title={The PASCAL Visual Object Classes {(VOC)} Challenge},
      author={Mark Everingham and Luc Van Gool and Christopher K. I. Williams and John Winn and Andrew Zisserman},
      year={2010},
      eprint={0909.5206},
      archivePrefix={arXiv},
      primaryClass={cs.CV}
}

\end{document}